\documentclass[aps,prd,preprint,longbibliography]{revtex4-2}
\usepackage{units}
\usepackage{amsmath,amssymb,amsthm,amstext,amsfonts}
\usepackage{graphicx}
\usepackage[pdftex,plainpages=false,pdfpagelabels]{hyperref}
\usepackage{subfigure}

\begin{document}

\title{A new bound on Lorentz violation based on the absence of vacuum Cherenkov radiation in ultra-high energy air showers}

\author{Fabian Duenkel}
\affiliation{Department Physik, Universit\"at Siegen, 57068 Siegen, Germany}
\author{Marcus Niechciol}
\affiliation{Department Physik, Universit\"at Siegen, 57068 Siegen, Germany}
\author{Markus Risse}
\affiliation{Department Physik, Universit\"at Siegen, 57068 Siegen, Germany}

\begin{abstract} 
In extensive air showers induced by ultra-high energy (UHE) cosmic rays,
secondary particles are produced with energies far above those accessible by other means.
These extreme energies can be used to search for new physics. We study the effects of isotropic, nonbirefringent Lorentz violation in the photon sector.
In case of a photon velocity smaller than the maximum attainable velocity of standard Dirac fermions, vacuum Cherenkov radiation becomes possible.
Implementing this Lorentz-violating effect in air shower simulations, a significant reduction of the calculated average atmospheric depth of the shower maximum $\left<X_\text{max}\right>$ is obtained. 
Based on $\left<X_\text{max}\right>$ and its shower-to-shower fluctuations $\sigma(X_\text{max})$, a new bound on Lorentz violation is derived which improves the previous one by a factor of 2. 
This is the first such bound based on the absence of vacuum Cherenkov radiation from fundamental particles (electrons and positrons) in air showers.
Options for further improvements are discussed.

\end{abstract}

\maketitle
\newpage

\section{Introduction}
Deviations from exact Lorentz symmetry can occur
according to present works towards a fundamental theory of particle physics and gravitation (for reviews on quantum gravity phenomenology, see, e.g.,~\cite{liberati09a,Addazi:2021xuf}).
The extreme energies of cosmic rays and gamma rays can be exploited
to probe various effects of Lorentz violation (LV),
and some of the best limits on LV were obtained this way \cite{Addazi:2021xuf,kostelecky11a,klinkhamer08a,klinkhamer08b,klinkhamer08c,klinkhamer17,duenkel2019,Risse:2022unt}.

Here, we focus on isotropic nonbirefringent LV in the photon sector.
We now specialize to the case of a photon velocity smaller than the maximum attainable velocity of standard Dirac fermions~\cite{kostelecky02a}
which allows vacuum Cherenkov (VCh) radiation as a new process.
The impact of this type of LV on the development of extensive air showers initiated by cosmic rays is investigated and exploited, 
with a focus on ultra-high energies (UHE) above $\unit[1]{EeV} = \unit[10^{18}]{eV}$.

The theory background of LV in the context of this study and relevant aspects of the previous studies are briefly summarized in Sec.~\ref{sec:theory}. 
The current analysis is presented in Sec.~\ref{sec:analysis}, including the methodology to compare simulations and data in more than one observable and the result after application. Sec.~\ref{sec:discussion} contains a discussion and a brief outlook.

\section{Theory background and previous bounds}
\label{sec:theory}

As in previous works \cite{klinkhamer17,duenkel2019}, we use a fairly simple extension of standard quantum electrodynamics
(QED). A single term is added to the Langrange density which breaks Lorentz invariance but preserves CPT and gauge invariance \cite{chadha83a,kostelecky02a}:
\begin{equation}
\begin{split}
\mathcal{L} = &\underbrace{-\frac{1}{4}F^{\mu\nu}F_{\mu\nu} +
  \overline{\psi}\left[\gamma^\mu(i\partial_\mu-eA_\mu)-m\right]\psi}_{\text{standard
  QED}}\\
&\underbrace{-\frac{1}{4}(k_F)_{\mu\nu\rho\sigma}F^{\mu\nu}F^{\rho\sigma}}_{\text{CPT-even
LV term}}.
\end{split}
\label{eq:lv_lagrangian}
\end{equation}
Natural units ($\hbar =  c = 1$) and the Minkowski metric $g_{\mu\nu}(x) = \eta_{\mu\nu} =   [\text{diag}(+1,-1,-1,-1)]_{\mu\nu}$ are used here.
The added (observer) tensor $(k_F)_{\mu\nu\rho\sigma}$ consists of 20 independent components. Ten of these produce birefringence, eight lead to direction-dependent modifications of photon propagation,
and one corresponds to an unobservable double trace that changes the normalization of the photon field. The remaining component causes an isotropic modification of the photon propagation.
Thus, isotropic nonbirefringent LV in the photon sector is controlled by a single dimensionless parameter $\kappa$ which is related to the tensor-valued background field $k_F$, that is fixed with respect to particle Lorentz transformations, in Eq.~(\ref{eq:lv_lagrangian}) by \cite{diaz15a}
\begin{equation}
{(k_F) ^{\lambda}}_{\mu\lambda\nu} = \frac{\kappa}{2}\left[\text{diag}(3,1,1,1)\right]_{\mu\nu}.
\label{eq:kF}
\end{equation}
The phase velocity of the photon is given by
\begin{equation}
v_\text{ph} = \frac{\omega}{|\vec{k}|} =
\sqrt{\frac{1-\kappa}{1+\kappa}}\ c.
\label{eq:phasevelocity}
\end{equation}
In physical terms, the velocity $c$, added in Eq.~(\ref{eq:phasevelocity}) for clarity, despite working in natural units, corresponds to the maximum attainable velocity of the massive Dirac fermion in Eq.~(\ref{eq:lv_lagrangian}), whereas the phase velocity $v_\text{ph}$ of the photon is smaller (larger) than $c$ for positive (negative) values of $\kappa$. We note that microscopic models exist for both positive ~\cite{Klinkhamer:2010zs,Bernadotte:2006ya} 
and negative~\cite{Klinkhamer:2011ez} values of $\kappa$
and continue to examine how the physics would be if this was realized in nature.

For non-zero values of $\kappa$, certain processes which are forbidden in the conventional, Lorentz-invariant theory become allowed. 
In a previous work \cite{duenkel2019}, we investigated the case $\kappa < 0$, which allows photons to decay. Applied to air shower simulations, this LV effect led to a reduction of the average atmospheric depth of the shower maximum $\left<X_\text{max}\right>$, and an improved bound on the negative $\kappa$ could be determined.
Here, we now focus on the case $\kappa > 0$. In this case, charged particles above the energy threshold
\begin{equation}
E^\text{th}_\text{VCh}(\kappa) = m\,\sqrt{\frac{1+\kappa}{2\kappa}} \simeq \frac{m}{\sqrt{2\kappa}},
\label{eq:VChthreshold}
\end{equation}
emit VCh radiation, where $m$ is the rest mass of the corresponding particle. Particles with an energy above this threshold continuously radiate VCh photons with an emission rate $\Gamma$ described by \cite{diaz15a}
\begin{equation}
\frac{d\Gamma}{d\omega}=\frac{\alpha Z^2}{E\sqrt{E^2-m^2}}\left[{\frac{2\kappa E}{1-\kappa^2}\left({E-\omega}\right)-\frac{m^2}{1-\kappa}+\frac{\kappa}{(1-\kappa^2)(1-\kappa)}\omega^2}\right]
\label{eq:VChemissionrate}
\end{equation}
where $Z$ is the particle charge in units of the proton charge $e$, $\alpha$ is the fine structure constant and $\omega$ is the photon energy. The energy of the radiated photon is cut off at an upper bound of 
\begin{equation}
\omega_{\text{max}} = \left({\frac{1-\kappa}{\kappa}}\right)\left[{\sqrt{\frac{1+\kappa}{1-\kappa}}\sqrt{E^2-m^2}-E}\right] \text{.}
\label{eq:VChcutoff}
\end{equation}

\begin{figure}[tp]
\centering
\includegraphics[width=0.75\textwidth,]{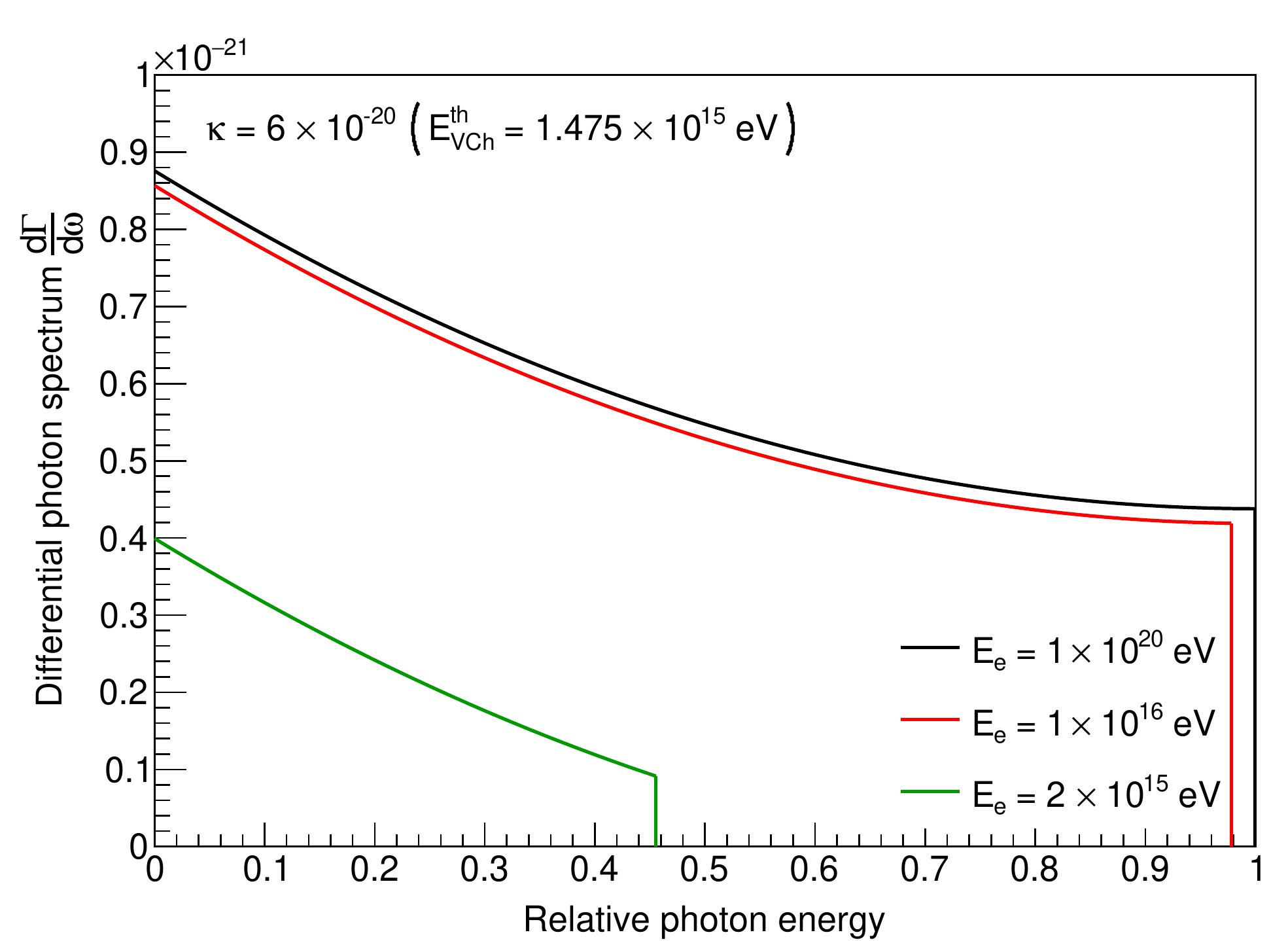}
\caption{Emission rates of VCh photons inheriting a relative energy fraction from a primary electron for $\kappa = 6 \times 10^{-20}$ and different electron energies.}
\label{fig:VCh_relphotonenergy} 
\end{figure}

The VCh interaction length drops to less than a centimeter right above the threshold \cite{klinkhamer08c,diaz15a},
resembling a quasi-instantaneous radiation of photons. These VCh photons can inherit a significant share of the energy of the primary particle. In Fig.~\ref{fig:VCh_relphotonenergy}, the relative energy of the emitted VCh photon is displayed for different electron energies and a given value of $\kappa$. The distribution is fairly flat, with a moderate enhancement towards small energy fractions. Unless just above threshold energy, photon energy fractions up to almost 100\% are possible. 

Due to the very efficient energy loss, above-threshold particles from astrophysical sources are not able to reach Earth.
Therefore, terrestrial observations of cosmic rays with energies of the order $\unit[100]{EeV}$
were used to impose a limit of~\cite{klinkhamer08b,klinkhamer08c}
\begin{equation}
\kappa < 6 \times 10^{-20} ~~~  \text{($\unit[98]{\%}$ CL)}~.
\label{eq:limitold}
\end{equation}
For this limit, the primary particle was assumed to be an iron nucleus, which can be regarded a conservative assumption.
Observations of higher-energy cosmic rays or a restriction of the primary type to nuclei lighter than iron would be needed to improve this limit.

Here, we explore a different approach and investigate whether VCh radiation, allowed for $\kappa > 0$, would impact the air shower development. This complements the recent study on the impact of photon decay, allowed for $\kappa < 0$, on the shower development~\cite{duenkel2019}.

In air showers initiated by UHE hadrons in the Earth's atmosphere, high-energy electrons are expected to be produced as secondary particles.
In the first interaction of the primary hadron with an atmospheric nucleus, mostly charged and neutral pions are produced.
The charged pions further interact with particles from the atmosphere, producing more pions, while the neutral pions,
in standard physics, rapidly decay into pairs of photons, which in turn trigger electromagnetic sub-showers.
For cosmic-ray primaries with an energy per nucleon $E/A >$~1~EeV, secondary electrons (and positrons) can be produced up to energies $>$~0.1~EeV.
Especially in the start-up phase of the air shower, where the energy of the secondary particles is very high, a modification of
the particles due to LV  can drastically modify the overall development of the air shower (examples for the case of $\kappa < 0$ can be seen in~\cite{klinkhamer17,diaz16a}).

\begin{figure}[p]
  \centering
  \subfigure[]{\label{fig:xmax_hadrons}\includegraphics[width=0.75\textwidth,]{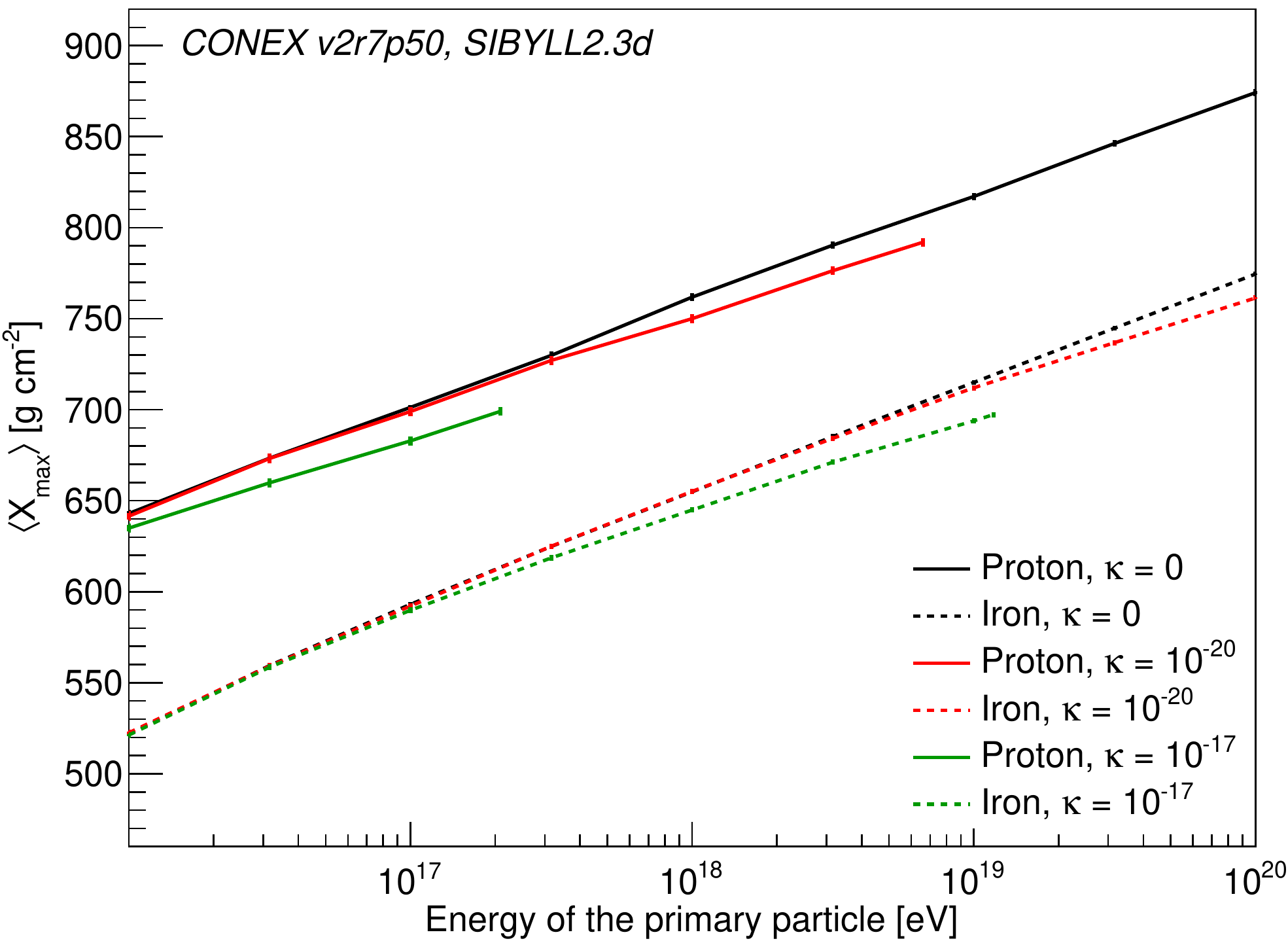}}\\
  \subfigure[]{\label{fig:sigma_hadrons}\includegraphics[width=0.75\textwidth,]{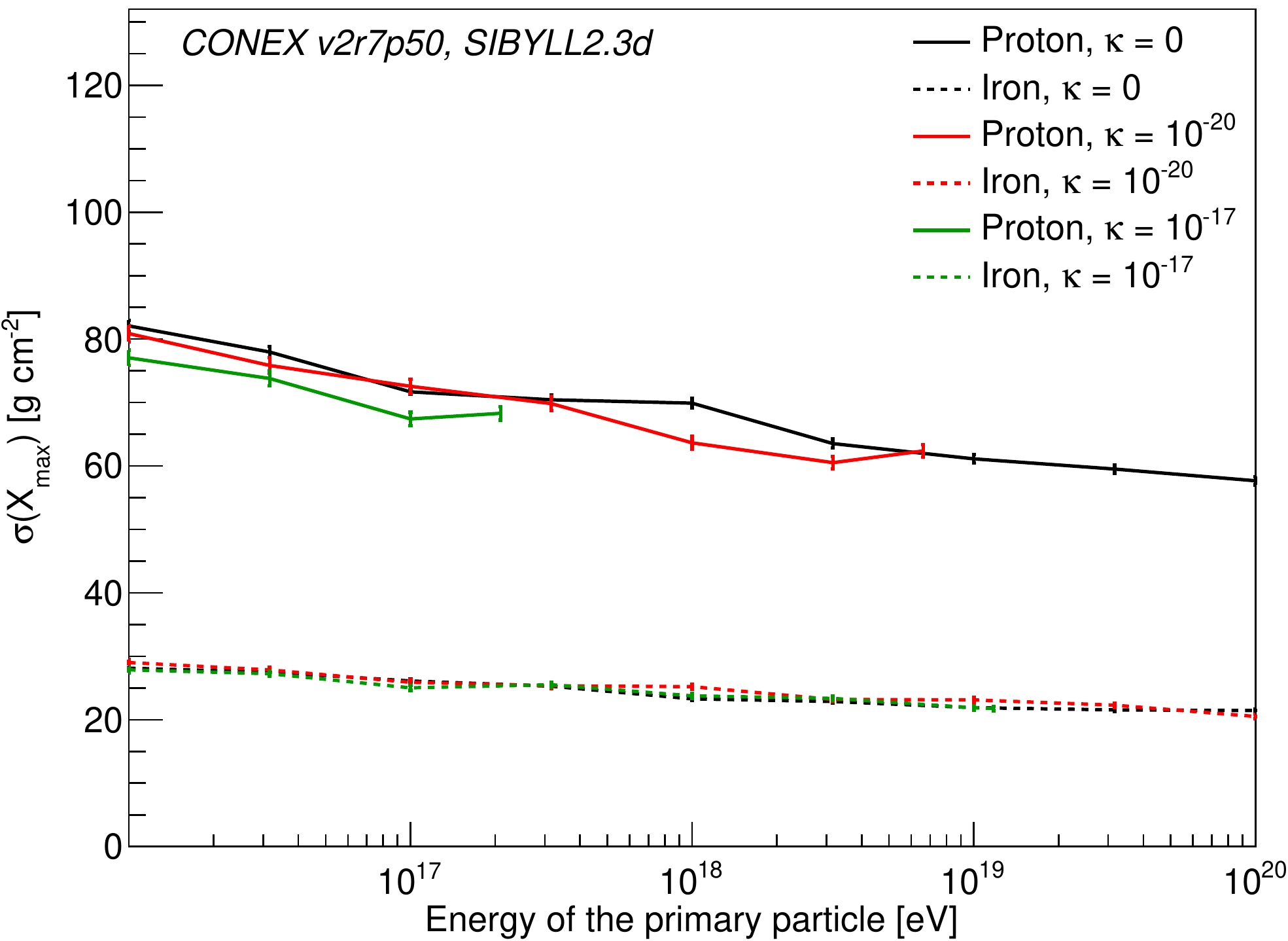}}\\
\caption{The simulated values of $\left<X_\text{max}\right>$ and $\sigma(X_\text{max})$ as a function of the primary energy for primary protons and iron nuclei for the absence of LV ($\kappa = 0$) and for different values of $\kappa$. Lines are only plotted up to the maximum energy given by the onset of VCh radiation, see Eq.~(\ref{eq:VChthreshold}).}
\label{fig:xmax_sigma_hadrons} 
\end{figure}

Overall, shorter showers could be expected in the case of VCh radiation, since the highest energy electrons in the startup-phase of the shower immediately emit VCh photons until the VCh threshold is reached. This effectively replaces those electrons with multiple photons of lower energy plus an electron below the energy threshold $E^\text{th}_\text{VCh}(\kappa)$. Those subshowers are then starting at lower energies and higher altitude than the ones the electron would have initiated without VCh radiation and, thus, typically evolve faster, resulting in a smaller value of $\left<X_\text{max}\right>$ for the entire shower.

A similar effect on $\left<X_\text{max}\right>$ resulting from $\kappa < 0$ and the related consequence of photon decay was already used to set stringent bounds on negative values of $\kappa$ in \cite{duenkel2019}. We adapted this method to apply it to $\kappa > 0$, as shown in the next section.

\section{Analysis}
\label{sec:analysis}

To analyze the potential impact of LV on the development of air showers, a full Monte Carlo (MC) approach as in~\cite{klinkhamer17,duenkel2019} is used.
We modified the MC code CONEX~\cite{bergmann07a,pierog06a} to include VCh radiation of electrons and positrons (simply termed ``electrons'' in the following). We focus for the time being on these particles, since they are the lightest charged particles (i.e., smallest $E^\text{th}_\text{VCh}$ for a given $\kappa$) and they dominate the longitudinal shower profile. 
Hadronic interactions are simulated with
EPOS LHC~\cite{pierog15a}, QGSJET-II-04~\cite{ostapchenko11a} and SIBYLL~2.3d~\cite{riehn20a} using CONEX v2r7p50.
The default settings provided by the CONEX code are used for all settings not mentioned here.
In the following analysis, results gained from simulations using SIBYLL~2.3d are presented where not stated otherwise, as this model gives the most conservative results.

The impact of LV and the resulting VCh radiation on $\left<X_\text{max}\right>$ and its fluctuations $\sigma(X_\text{max})$ is presented in Fig.~\ref{fig:xmax_sigma_hadrons}.
A reduction of $\left<X_\text{max}\right>$ for primary energies well above the VCh radiation threshold can be seen. The reduction is larger with growing $ |\kappa |$ and growing $E/A$. For instance, $\left<X_\text{max}\right>$ is reduced by $\sim$15~g~cm$^{-2}$ for 3~EeV proton showers with $\kappa = 10^{-20}$.
Depending on $\kappa$, a maximum primary energy exists above which the primary particle would itself emit VCh radiation and not be able to reach earth at all: thus, the simulated lines in Fig.~\ref{fig:xmax_sigma_hadrons} are only plotted up to this maximum energy. We return to this effect below.

The reduction of $\left<X_\text{max}\right>$ is smaller for positive $\kappa$ values than for negative values of the same magnitude. The main reason for this difference is the fact that subshowers initiated by photons are typically deeper than those by electrons of same initial energy. For negative $\kappa$, photons produce electrons/positrons, while for positive $\kappa$ the opposite is the case (electrons/positrons produce photons), resulting in a less pronounced reduction of $\left<X_\text{max}\right>$ for positive $\kappa$. 

Concerning $\sigma(X_\text{max})$, no strong effect can be noted, as in the case of negative $\kappa$~\cite{duenkel2019}. This is expected since a main effect on shower fluctuations results from the depth of the first interaction, i.e., before the appearance of VCh radiation.

For the further simulations, we have chosen five elements as representatives of their respective mass ranges in order to cover the range of possible primary compositions: protons (mass number $A = 1$), helium ($A = 4$), nitrogen ($A$ = 14), silicon ($A = 28$) and iron ($A = 56$). 
The simulated showers were combined using different weights to obtain sets of air showers induced by different primary hadron combinations.
The chosen stepsize between different relative contributions of individual hadrons was $\unit[2]{\%}$.

\begin{figure}[tp]
\centering
\includegraphics[width=0.75\textwidth,]{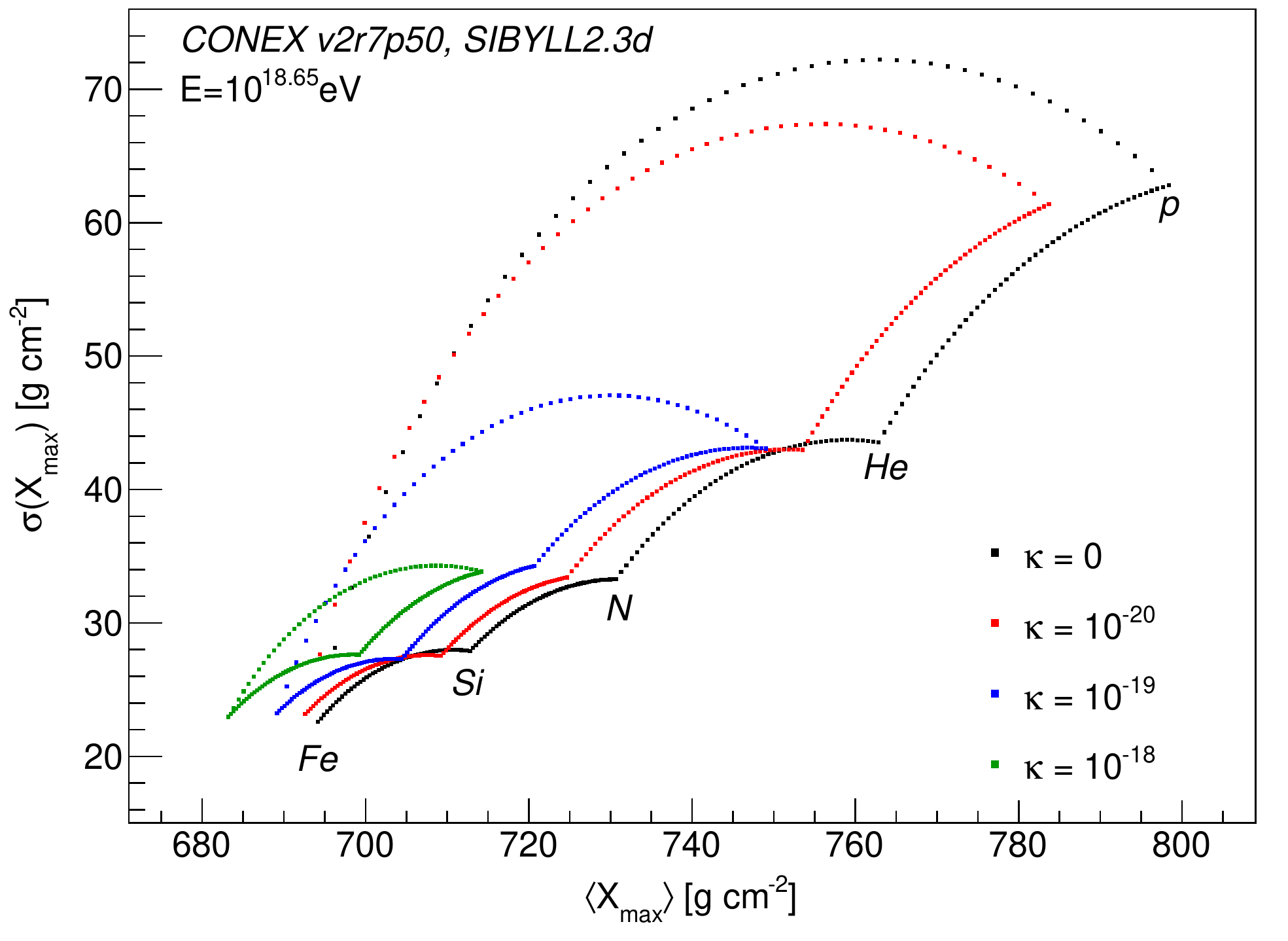}
\caption{Comparison of combinations of $\left<X_\text{max}\right>$ and $\sigma(X_\text{max})$ derived by simulations, excluding combinations with particles above the VCh radiation threshold, for different values of $\kappa$ and a primary energy of $\unit[10^{18.65}]{eV}$. The ``tips'' of the umbrella refer to pure beams with primaries as indicated (proton, helium, nitrogen, silicon, iron). For a given $\kappa$, the corresponding ``umbrella'' covers all values allowed by arbitrary combinations of these primaries. With increasing $\kappa$, a reduction of $\left<X_\text{max}\right>$ can be seen as well as a successive removal of light primaries if above VCh radiation threshold.}
\label{fig:umbrella_restrictedcomps} 
\end{figure}

An example of the possible range of combinations of $\left<X_\text{max}\right>$ and $\sigma(X_\text{max})$ for a fixed value of primary hadron energy and $\kappa$ is displayed in Fig.~\ref{fig:umbrella_restrictedcomps}.
The distinctive ``umbrella''-like shape is visible (cf.\ also~\cite{duenkel2019}). 
The reduction of $\left<X_\text{max}\right>$ with increasing value of $\kappa$ can be seen.
For any composition, the $\left<X_\text{max}\right>$ value corresponds to the weighted mean of the $\left<X_\text{max}\right>$ values of all components, while the $\sigma(X_\text{max})$ value of a set of showers with different primary hadrons is always greater than the weighted mean of the composites. 
This effect is due to the size of the shower-to-shower fluctuations increasing once showers induced by different particles with different mean shower depths are combined. 

VCh radiation might not only affect secondary electrons in the shower, but also the primary particles, making it impossible for lighter particles to arrive at higher energies for sufficiently high $\kappa$. This restriction on possible primary masses can also be seen in Fig.~\ref{fig:umbrella_restrictedcomps} in the form of smaller umbrellas at higher $\kappa$ values resulting from the successive removal of light primaries with increasing $\kappa$. The lower limit of the allowed primary particle masses is, following  Eq.~(\ref{eq:VChthreshold}), 
\begin{equation}
m \geq E_0 \,\sqrt{2\kappa},
\label{eq:MassVChthreshold}
\end{equation}
where $E_0$ is the energy of the primary particle.

\begin{figure}[tp]
\centering
\includegraphics[width=0.75\textwidth,]{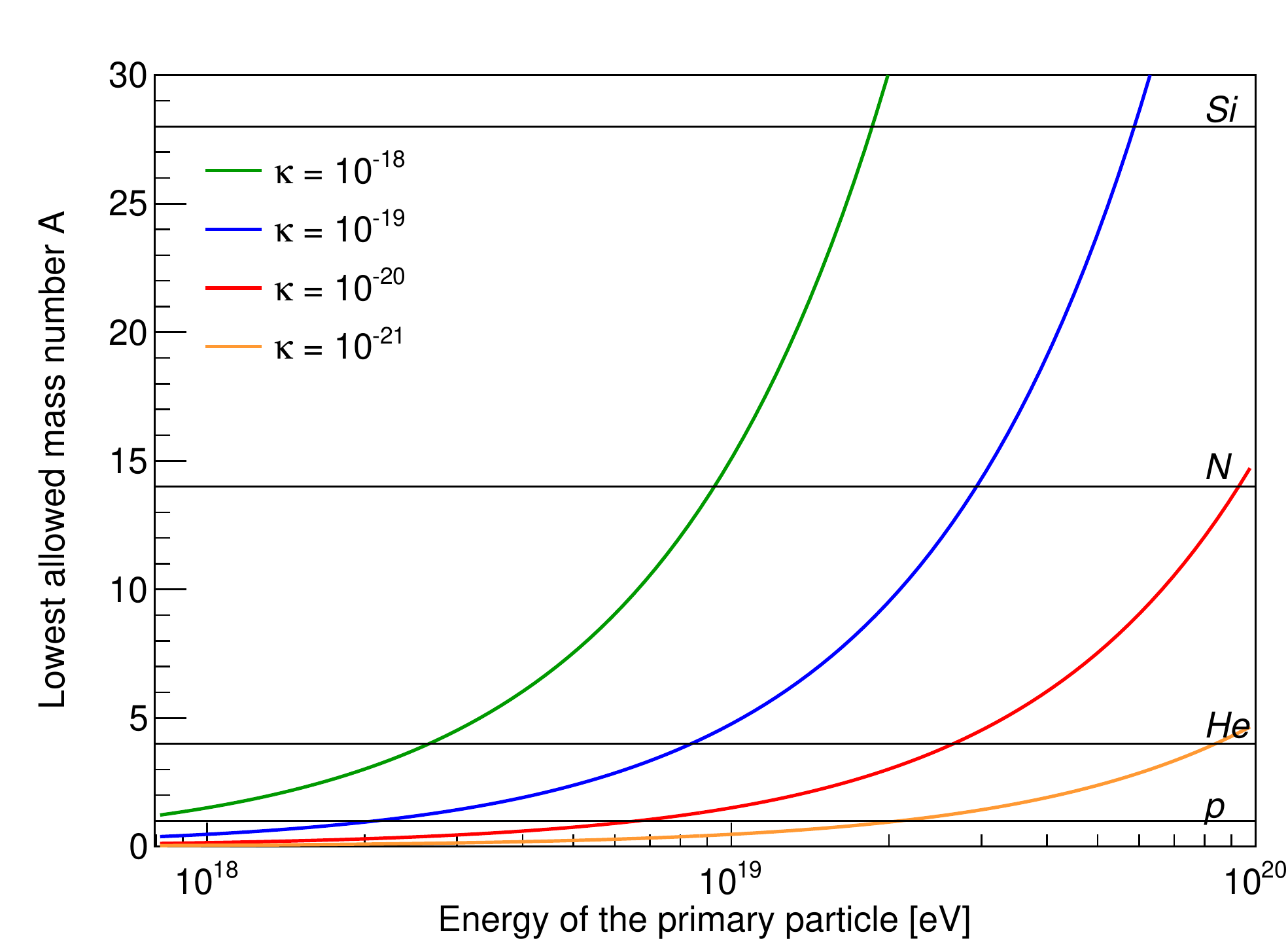}
\caption{Mass threshold (in terms of mass number $A$) related to VCh radiation for different values of $\kappa$ as a function of primary energy, cf.~Eq.~(\ref{eq:MassVChthreshold}). For a given $\kappa$, only primaries above the line can reach the earth.}
\label{fig:VCh_minimalmasses} 
\end{figure}

In Fig.~\ref{fig:VCh_minimalmasses} an illustration of this limit on the lowest possible particle mass is presented.
This restriction shrinks the range of possible combinations for larger $\kappa$ values, which can be seen in Fig.~\ref{fig:umbrella_restrictedcomps} as the tips of the umbrella plot representing lighter masses successively disappearing for higher values of $\kappa$.

\begin{figure}[tp]
\centering
\includegraphics[width=0.75\textwidth,]{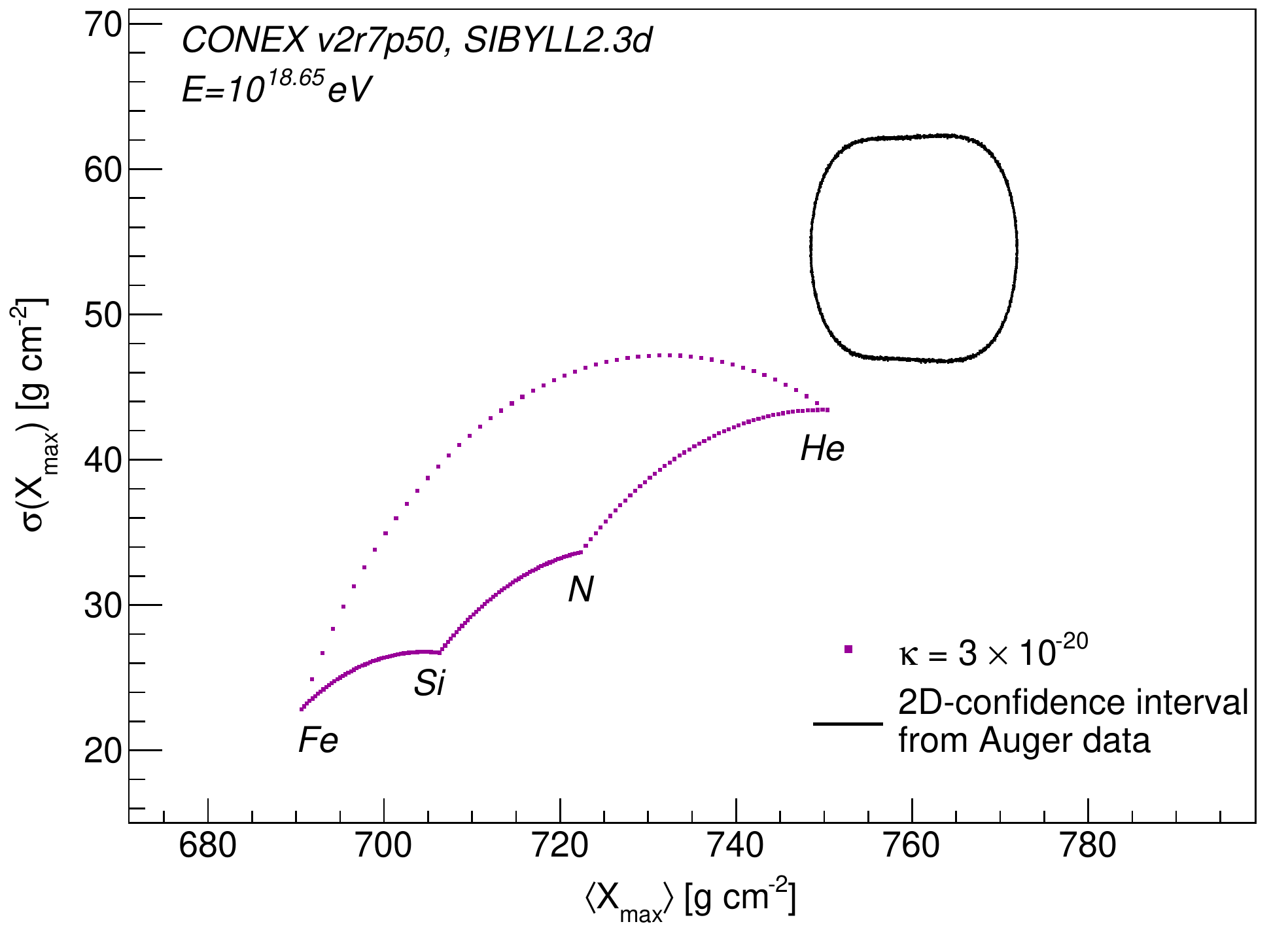}
\caption{Comparison of the combinations of $\left<X_\text{max}\right>$ and $\sigma(X_\text{max})$ derived by simulations which incorporate LV to the 2D confidence interval ($\unit[98]{\%}$ CL) given by the measurements of the Pierre Auger Observatory~\cite{auger19} for the critical value of $\kappa = 3 \times 10^{-20}$ and a primary particle energy of $\unit[10^{18.65}]{eV}$. The umbrella shape covers all combinations of iron, silicon, nitrogen and helium primary particles, since protons are excluded due to VCh radation for these values of energy and $\kappa$.}
\label{fig:umbrella_critkappa} 
\end{figure}

The sets of simulated possible values obtained this way are then compared to the measurements taken by the Pierre Auger Observatory~\cite{auger19} following our method developed previously~\cite{duenkel2019}.
To accomplish a simultaneous comparison of both observables $\left<X_\text{max}\right>$ and $\sigma(X_\text{max})$, in each energy bin a two-dimensional confidence interval was used (at a confidence level of $\unit[98]{\%}$). 
For this, the statistical and systematic uncertainties of the $\left<X_\text{max}\right>$ and $\sigma(X_\text{max})$ observations are approximated by Gaussian distributions (statistical) and uniform distributions (systematic)
and a contour line encompassing $\unit[98]{\%}$ of the distribution is drawn. The comparison is performed between all possible combinations of $\left<X_\text{max}\right>$ and $\sigma(X_\text{max})$ covered by the LV simulations and the Auger measurements. 
An illustration of such a comparison for the later derived limit on $\kappa$ can be seen in Fig.~\ref{fig:umbrella_critkappa}. 

An overlap between two areas (simulated vs.\ observed) in the figure would show that there are primary hadron combinations which fit the Auger measurements.
Conversely, if for a specific value of $\kappa$ there is an energy at which no primary hadron combination fits the Auger measurements, it means this $\kappa$ does not fit the measurements and can thus be excluded.
This procedure is repeated, scanning over $\kappa$ and the data energy bins, until a value $\kappa_{\text{crit}}$ is found where all $\kappa>\kappa_{\text{crit}}$ do not fit the measurements while all $\kappa<\kappa_{\text{crit}}$ do. 
We excluded, for the time being, the two highest-energy bins (above $\unit[10^{19.50}]{eV}$) due to the comparably small statistics.
Comparing the simulations to data, this results in the new limit of

\begin{equation}
\kappa < \kappa_{\text{crit}} = 3 \times 10^{-20} ~~~  \text{($\unit[98]{\%}$ CL)}~.
\label{eq:newlimitcommonIsotopes}
\end{equation}

This limit refers to the energy bin $\unit[10^{18.6}]{eV}$ to $\unit[10^{18.7}]{eV}$ and improves the previous bound of Eq.~(\ref{eq:limitold}) by a factor of 2. It is obtained by both hadronic interaction models SIBYLL~2.3d and EPOS LHC. The model QGSJET-II-04 produces showers which are not able to reproduce the data in a self-consistent way, even for $\kappa = 0$ (no LV), due to much shallower showers.
This known fact (see e.g.~\cite{auger19}) indicates shortcomings in this specific hadronic interaction model.

\section{Discussion}
\label{sec:discussion}

While electrons make up the overwhelming majority of particles in air showers which would undergo VCh radiation, in principle also other charged particles could be above the VCh threshold for sufficiently high $E/m$. Both pions and muons are relatively light particles, which can appear in the early stages of the air-shower development. Muons are structureless charged fermions and emit VCh radiation if above-threshold. For pions, VCh radiation can come from the partons, similar to the case of neutrons considered in \cite{diaz15a}. For pions and muons inheriting more than $\sim$10\% of the energy per nucleon in the first interaction, VCh radiation can happen, although this is a relatively rare case. For completeness, VCh radiation of pions and muons had also been implemented. In a dedicated comparison switching the effect for pions and muons on or off, no significant impact on the longitudinal development is found, with differences in  $\left<X_\text{max}\right>$ and $\sigma(X_\text{max})$ in the relevant energy range below 1~g~cm$^{-2}$.

The new limit (Eq.~\ref{eq:newlimitcommonIsotopes}) is quite robust against the specific energy bin chosen. The bound from other energy bins is only slightly larger ($\kappa < 4 \times 10^{-20}$).

The limit is driven by the energy bin in the energy range from $\unit[10^{18.6}]{eV}$ to $\unit[10^{18.7}]{eV}$ with a mean energy of $\unit[10^{18.65}]{eV}$. At this combination of energy and $\kappa$, the occurence of protons is excluded due to VCh, leading to helium ($A$ = 4) being the lightest possible particle. Deuterium and helium-3 were excluded in this analysis due to them making up only a low portion of cosmic rays \cite{Adriani:2016}. Allowing a portion of these isotopes of up to $40\%$ does not change the derived limit. For the somewhat extreme assumption that any stable isotope could contribute with an arbitrary proportion, a limit of $\kappa < 6 \times 10^{-20}$ is obtained in the energy bin around $\unit[10^{19.05}]{eV}$ if the composition consisted of $100 \%$ lithium-6.

To conclude, a new method has been developed to test the case of LV for $\kappa > 0$. Compared to the previous work \cite{klinkhamer08b,klinkhamer08c}, the level of the limit is confirmed and improved by a factor of 2. In contrast and complementing to the previous bound, the present one is based on air shower quantities (instead of primary particles) and on fundamental particles, electrons and positrons (instead of nuclear primaries).

Future improvements may be related to a reduction of the data uncertainty, e.g.\ by increased statistics particularly at the highest energies. For instance, uncertainties reduced by a factor of 2 could lead to an improvement of the limits by a factor of $\sim$3. Also, constraints on the allowed range of primary compositions to be considered may help to increase the sensitivity. As an example, constraints on the purity of the primary beam (cf.~\cite{PierreAuger:2016qzj}) would remove the tips of the umbrellas and, thus, potentially exclude the LV hypothesis for values of $\kappa$ that are presently still allowed.

In this work, we restricted ourselves to the impact of LV on the longitudinal shower development. A study of the muon content and its suitability to test LV is left for a future work.

\subsection*{Acknowledgments}
The many fruitful discussions with Frans R. Klinkhamer are greatly appreciated.
We thank T. Pierog for his help in modifying the CONEX source code.
This work was funded by the German Research Foundation (DFG project 408049454).

\bibliography{bibl.bib}

\end{document}